\documentclass[10pt, conference]{IEEEtran}
\usepackage{fancyhdr}

\usepackage{cite}
\usepackage{amsmath,bm,amssymb,amsfonts}
\usepackage{algorithm}
\usepackage{algpseudocode}
\usepackage{graphicx}
\usepackage{textcomp}
\usepackage[table]{xcolor}
\usepackage{booktabs, multirow, multicol}
\usepackage{threeparttable}
\usepackage{tablefootnote}
\usepackage{amssymb}%
\usepackage{pifont}%
\usepackage{makecell}
\usepackage{arydshln}
\usepackage[hidelinks,breaklinks=true]{hyperref}
\usepackage[capitalise,noabbrev]{cleveref}
\usepackage{tikz}
\usepackage{soul}
\usetikzlibrary{decorations.pathreplacing, arrows.meta, arrows}
\usepackage{xcolor}
\usepackage{microtype}
\newcounter{algsubstate}
\renewcommand{\thealgsubstate}{\alph{algsubstate}}

\newcommand{\bzero}{\mathbf{0}}

\newcommand{\br}{\boldsymbol{\textbf{r}}}

\newcommand{\bx}{\boldsymbol{\textbf{x}}}

\newcommand{\bB}{\boldsymbol{\textbf{B}}}

\newcommand{\bH}{\boldsymbol{\textbf{H}}}

\newcommand{\bM}{\boldsymbol{\textbf{M}}}

\newcommand{\bP}{\boldsymbol{P}}
\newcommand{\bQ}{\boldsymbol{Q}}
\newcommand{\bR}{\boldsymbol{\textbf{R}}}

\newcommand{\bX}{\boldsymbol{X}}
\newcommand{\bY}{\boldsymbol{Y}}

\newcommand{\bLam}{\boldsymbol{\Lambda}}
\newcommand{\bDelta}{\boldsymbol{\Delta}}
\newcommand{\bD}{\boldsymbol{D}}
\newcommand{\bZ}{\boldsymbol{Z}}

\newcommand{\bsigma}{\boldsymbol{\sigma}}
\definecolor{hellgruen}{rgb}{0.2,0.7,0.2}
\newcommand{\cb}[1]{{\color{blue}#1}}

\newcommand{\cn}{\color{black}}

\definecolor{c1}{RGB}{160,230,50}
\definecolor{c2}{RGB}{238,93,8}
\definecolor{c3}{RGB}{226,24,110}
\definecolor{c4}{RGB}{71,31,236}
\newcolumntype{P}[1]{>{\centering\arraybackslash\hspace{0pt}}p{#1}}
\newcolumntype{M}[1]{>{\centering\arraybackslash}m{#1}}
\makeatletter
\newcommand*{\rom}[1]{\expandafter\@slowromancap\romannumeral #1@}
\newcommand{\equalcontrib}{\IEEEauthorrefmark{\@fnsymbol{footnote}}}
\makeatother

\def\BibTeX{{\rm B\kern-.05em{\sc i\kern-.025em b}\kern-.08em
    T\kern-.1667em\lower.7ex\hbox{E}\kern-.125emX}}

\begin{document}

\title{Towards Exascale Fully Relativistic Pseudopotential Density Functional Theory Calculations with Mixed-Precision Computation and Compressed Communication Using Residual-Based Subspace Iteration}

\author{
  \IEEEauthorblockN{
    Nikhil Kodali$^*$,
    Gourab Panigrahi$^*\IEEEauthorrefmark{6}$,
    Nishant Gupta$^*\IEEEauthorrefmark{6}$,
    Sundaresan Gourishanker$^*$,
    Kartick Ramakrishnan$^*$,\\
    Rudra Panch$^*$,
    Sambit Das$^\dagger$,
    Vishwas Rao$^\ddagger$,
    Phani Motamarri$^{*\P}$
  }
  \IEEEauthorblockA{$^*$Department of Computational and Data Sciences,
    Indian Institute of Science, Bangalore, India\\
    $^\dagger$Department of Mechanical Engineering,
    University of Michigan, Ann Arbor, USA\\
    $^\ddagger$Argonne National Laboratory, Lemont, IL, USA
  }
}

\maketitle
\thispagestyle{fancy}
\lhead{}
\rhead{}
\chead{}
\lfoot{\footnotesize{SC26, November 15-20, 2026, Chicago, Illinois, USA\newline 979-8-3195-4789-7/26/\$31.00 \copyright 2026 IEEE}}
\rfoot{}
\cfoot{}
\renewcommand{\headrulewidth}{0pt}
\renewcommand{\footrulewidth}{0pt}
\begingroup
\renewcommand{\thefootnote}{\ensuremath{\|}}
\footnotetext{Both authors contributed equally to this research}
\endgroup
\begingroup
\renewcommand{\thefootnote}{\P}
\footnotetext{Corresponding author: \texttt{phanim@iisc.ac.in}}
\endgroup

\begin{abstract}
Materials exhibiting noncollinear magnetism or strong spin-orbit-coupling underpin many spintronic and topological applications, but their simulations require complex two-component spinors and costs substantially more than scalar density functional theory (DFT). We present a GPU-centric exascale finite-element DFT framework with noncollinear magnetism and spin-orbit-coupling, combining (i) adaptive higher-order finite-element discretization, (ii) a matrix-free Poisson solver, (iii) residual-based Chebyshev filtered subspace iteration (R-ChFSI) for sparse generalized eigenproblems, (iv) R-ChFSI-enabled mixed precision computation with block floating-point compressed MPI communication, and (v) communication-efficient band partitioning. R-ChFSI permits inexact matrix-multivector products and $4\times$--$6.4\times$ compression of communicated data relative to FP64 while preserving double-precision robustness and reducing computation and data movement. Results demonstrate strong scaling on Aurora and Frontier, with up to $3.4\times$ faster Chebyshev filtering. On Aurora, total self-consistent-field (SCF) solve wall time reduces by $2\times$. An 800-node (9,600-GPU) calculation demonstrates fully relativistic pseudopotential DFT for $\sim$80,000 electrons in under 7 minutes per SCF iteration.
\end{abstract}

\IEEEpeerreviewmaketitle

\section{Introduction}

Understanding the quantum mechanical behavior of electrons in molecules and materials lies at the heart of modern computational science. Electronic structure methods, ranging from configuration interaction (CI) and quantum Monte Carlo (QMC) to density functional theory (DFT), have become indispensable tools for predicting and interpreting the properties of complex molecular and materials systems, and have underpinned several Gordon Bell Prize-winning~\cite{Gygi2006, Alvarez2008, Hasegawa2011, Ziogas2019, Das2023} and finalist~\cite{Fattebert2016, Das2019} works. Among these, DFT~\cite{Hohenberg1964, Kohn1965} stands out as the method of choice for large-scale quantum simulations, striking a compelling balance between computational efficiency and predictive accuracy. In particular, DFT offers a formally exact reduction of the otherwise intractable interacting many-electron Schr\"odinger equation to an equivalent problem of non-interacting electrons in an effective mean-field dependent on the electron density ($\rho(\br)$). In other words, conventional Kohn-Sham DFT calculations typically scale as $\mathcal{O}({N_{\mathrm{el}}}^3)$, rather than exponentially with the number of electrons $N_{\mathrm{el}}$. The development of DFT was recognized by the 1998 \emph{Nobel Prize} in Chemistry~\cite{DFTNobel}.

In many magnetic materials, the magnetization direction varies in space, giving rise to noncollinear (NC) magnetism. When such materials also contain heavy elements, spin-orbit coupling (SOC), the relativistic interaction between an electron's spin and orbital motion, becomes particularly important because it can couple magnetic moments to the crystal lattice and help stabilize noncollinear structures. In nonmagnetic materials, SOC can also lift certain band degeneracies and stabilize topological phases. Large-scale NC-SOC-DFT simulations are essential for understanding spin textures such as skyrmions and spin spirals, the nonlinear Hall effect in incommensurate Moir\'{e} lattices, anomalous transport in topological semimetals, and magnetocrystalline anisotropy in rare-earth permanent magnets with point defects. Predictive simulations of these phenomena are key to the rational design of spintronic devices including spin-transfer torque memories, racetrack storage, and spin-logic architectures, as well as topological qubits for fault-tolerant quantum computing and energy-efficient spin-based electronics that overcome the power limitations of conventional charge-based CMOS technology. However, generalizing Kohn-Sham DFT to incorporate noncollinear magnetism (NC) and SOC in simulations is far from trivial, as it requires complex two-component Kohn-Sham spinors that double both the Hamiltonian size and the number of eigenpairs to be
computed in complex arithmetic. This significantly increases the computational prefactor relative to standard DFT and drives the system to the asymptotic cubic-scaling regime at much smaller system sizes, limiting typical state-of-the-art NC-SOC-DFT calculations to ${\sim}500$--$2000$ electrons.

Efforts to circumvent the cubic-scaling bottleneck have focused on linear-scaling or reduced scaling approaches that exploit the nearsightedness of electronic 
interactions~\cite{Goedecker1999,Skylaris2005,Bowler2012,Wang2008,Eisenbach2009}, as well as stochastic DFT methods~\cite{Baer2013,Fabian2019}. However, these 
approaches are more challenging for metallic systems, where the density matrix decays more slowly, and the robustness of these techniques has not been broadly demonstrated for material systems exhibiting noncollinear magnetism and spin-orbit coupling. Consequently, a
systematically convergent and broadly applicable computational framework that reduces the prefactor for conducting large-scale NC-SOC-DFT on modern GPU-accelerated exascale architectures remains an open challenge. Compounding this, modern GPU architectures are increasingly optimized for low-precision arithmetic, offering substantially higher floating-point throughput at reduced precision compared to double precision. Yet scientific computing demands high accuracy, making it imperative to develop algorithms that exploit the performance gains of low-precision computation while preserving the robustness of double-precision results. %

Addressing the above challenges, we present a significant advance in the state of the art for 
NC-SOC-DFT calculations through a real-space computational framework that combines 
several algorithmic and HPC innovations applicable to metallic, semiconducting, and insulating 
systems. This is realized by bringing together the following ideas: (i) an 
efficient adaptive spatial discretization through higher-order finite elements 
(FE); (ii) a GPU-accelerated conjugate gradient solver using a novel matrix-free implementation strategy for
the Poisson problem in DFT electrostatics; (iii) residual-based Chebyshev filtered 
subspace iteration (R-ChFSI), tolerant to inexact matrix--multivector products, for
solving the resulting large, sparse generalized eigenproblem; (iv) mixed-precision 
computation and block floating-point compressed MPI communication, both enabled by R-ChFSI, preserving double-precision 
robustness while reducing compute and data movement costs; and (v) a 
communication-efficient band-partitioning algorithm to improve strong scalability 
on multinode GPU architectures at extreme scaling. The proposed framework enables NC-SOC-DFT 
calculations on systems with $\mathcal{O}(10^5)$ electrons, extending prior proof-of-concept NC-SOC-FE calculations~\cite{KodaliNC2025} at $\mathcal{O}(10^4)$ electrons and typical NC-SOC-DFT calculations at $\mathcal{O}(10^2)$--$\mathcal{O}(10^3)$ electrons. We implement
our framework in the electronic structure package DFT-FE~\cite{Das2019,Das2022} and evaluate it on the
exascale supercomputers Aurora (Intel GPUs) and Frontier (AMD GPUs), for both
semiconducting and metallic systems exhibiting noncollinear magnetism and 
spin-orbit coupling. In particular, we demonstrate our innovations on two 
scientifically important material classes: twisted bilayers of the heavy-element transition-metal dichalcogenide MoSe$_2$ (semiconducting) and the van der Waals system
Fe$_3$GeTe$_2$ (metallic), where large supercells and heavy atomic species make NC-SOC-DFT 
calculations particularly demanding. As demonstrated, the proposed strategies achieve substantial time-to-solution improvements over existing state-of-the-art baselines and excellent scalability on multinode GPU architectures, with HPC 
innovations broadly applicable to the wider computational chemistry and materials science communities.

\section{Background}
\subsection{NC-SOC-DFT Governing Equations}
Incorporating noncollinear magnetism into DFT requires
representing single-electron wavefunctions as two-component
complex spinors~\cite{Kubler1988,vonBarth1972}
\begin{equation}
  \Psi_n(\mathbf{r}) =
  \begin{pmatrix} \psi_n^{\uparrow}(\mathbf{r}) \\
                  \psi_n^{\downarrow}(\mathbf{r})
  \end{pmatrix}
  \quad \forall\, n = 1,\ldots,N_e,
\end{equation}
where $N_e$ is the number of computed spinors. The electron
charge density $\rho(\mathbf{r})$ and the magnetization
density $\mathbf{m}(\mathbf{r})$ are given by
\begin{equation}
  \rho(\mathbf{r}) = \sum_n f_n \Psi_n^\dagger(\mathbf{r})\Psi_n(\mathbf{r}),
  \quad
  \mathbf{m}(\mathbf{r}) = \sum_n f_n \Psi_n^\dagger(\mathbf{r})\,\vec{\bsigma}\,\Psi_n(\mathbf{r}),
\end{equation}
where $f_n\in[0,1]$ are the Fermi occupation numbers and $N_{\mathrm{el}}=\sum_{n=1}^{N_e}f_n$ is the number of electrons, while $\dagger$ denotes
the complex-conjugate transpose, and
$\vec{\bsigma} = \bsigma_x\hat{x} + \bsigma_y\hat{y} + \bsigma_z\hat{z}$
is the Pauli vector, with $\bsigma_x=\left(\begin{smallmatrix}0&1\\1&0\end{smallmatrix}\right)$, $\bsigma_y=\left(\begin{smallmatrix}0&-i\\i&0\end{smallmatrix}\right)$, and $\bsigma_z=\left(\begin{smallmatrix}1&0\\0&-1\end{smallmatrix}\right)$. The Kohn-Sham (KS) equations then
take the form of a nonlinear Hermitian eigenvalue
problem~\cite{Kubler1988,Kohn1965}
\begin{equation}
  H\Psi_n = \epsilon_n \Psi_n,
  \label{eq:ks_eigenproblem}
\end{equation}
to be solved for the $N_e$ spinors corresponding to the
$N_e$ smallest eigenvalues. The Hamiltonian operator can be decomposed
as $H = H_{\mathrm{loc}} + H_{\mathrm{nloc}}$, where
the local part is
\begin{equation}
  H_{\mathrm{loc}} =
  \left(-\tfrac{1}{2}\nabla^2 + V_{\mathrm{eff}}(\mathbf{r})\right)\mathbf{I}
  + \mathbf{B}_{\mathrm{xc}} \cdot \vec{\bsigma},
  \label{eq:hloc}
\end{equation}
with $\mathbf{I}$ the $2\times2$ identity matrix. The
effective potential and exchange-correlation magnetic
field are
\begin{equation*}
  V_{\mathrm{eff}} = V_{\mathrm{el}} + V_{\mathrm{xc}} +
  \left(V_{\mathrm{loc}} - V_{\mathrm{self}}\right),
  \;
  V_{\mathrm{xc}} = \frac{\delta E_{\mathrm{xc}}}{\delta\rho},
  \;
  \mathbf{B}_{\mathrm{xc}} = \frac{\delta E_{\mathrm{xc}}}{\delta\mathbf{m}}.
  \label{eq:veff}
\end{equation*}
Here, $E_{\mathrm{xc}}[\rho,\mathbf{m}]$ is the exchange-correlation energy functional, while $\delta/\delta\rho$ and $\delta/\delta\mathbf{m}$ denote functional derivatives. The local pseudopotential is given by $V_{\mathrm{loc}}(\mathbf{r},\mathbf{R})=\sum_a V_{\mathrm{loc}}^a(\mathbf{r}-\mathbf{R}_a)$, where $\mathbf{R}$ denotes the collection of atomic positions $\{\mathbf{R}_a\}$ with the sum running over atoms $a$, and each $V_{\mathrm{loc}}^a$ is specified by the pseudopotential data. The total nuclear self-interaction potential is $V_{\mathrm{self}}(\mathbf{r},\mathbf{R})=\sum_a V_{\mathrm{self}}^a(\mathbf{r},\mathbf{R}_a)$.
The total electrostatic potential $V_{\mathrm{el}}$ and
the nuclear self-interaction potential $V_{\mathrm{self}}^a$
are obtained by solving the Poisson problems~\cite{Motamarri2020,Das2022}
\begin{equation}
  -\nabla^2 V_{\mathrm{el}}(\mathbf{r}) = 4\pi\bigl(\rho(\mathbf{r}) + b(\mathbf{r},\mathbf{R})\bigr),
  \label{eq:poisson_el}
\end{equation}
\begin{equation}
  -\nabla^2 V_{\mathrm{self}}^a(\mathbf{r},\mathbf{R}_a) = 4\pi\, b^a(\mathbf{r} - \mathbf{R}_a),
  \label{eq:poisson_self}
\end{equation}
where $b(\mathbf{r},\mathbf{R})=\sum_a b^a(\mathbf{r}-\mathbf{R}_a)$
is the total smeared nuclear charge density. The
non-local part $H_{\mathrm{nloc}}$ encodes the
spin-orbit interaction via the separable optimized
norm-conserving Vanderbilt (ONCV)
pseudopotential~\cite{oncv2013,DalCorsoSOC2005} and
acts on a spinor as
\begin{equation}
  H_{\mathrm{nloc}}\Psi_n(\mathbf{r}) =
  \int V_{\mathrm{nloc}}(\mathbf{r},\mathbf{r}')\,
  \Psi_n(\mathbf{r}')\, d\mathbf{r}',
  \label{eq:hnloc}
\end{equation}
where $V_{\mathrm{nloc}}(\mathbf{r},\mathbf{r}')$ is
the non-local pseudopotential kernel whose $2\times2$
matrix structure in spin space captures the full
relativistic coupling between spin-orbit split
angular-momentum channels. We note that Equation~\eqref{eq:ks_eigenproblem} is a nonlinear eigenvalue problem solved as a sequence of linear eigenproblems using a self-consistent field (SCF) iteration. At each SCF iteration, the input charge and magnetization densities, $\rho$ and $\mathbf{m}$, determine $V_{\mathrm{eff}}$ and $\mathbf{B}_{\mathrm{xc}}$, the resulting Hamiltonian eigenproblem is solved, and updated $\rho$ and $\mathbf{m}$ are formed from the spinors. The densities are then mixed and the process is repeated until convergence.

\subsection{State of the Art}

Solving the DFT eigenproblem in practice requires a
suitable discretization of the governing equations. The
most widely used approach in solid-state calculations is
the plane-wave (PW)
basis~\cite{Kresse1996,Giannozzi2017}, which enjoys
spectral convergence but suffers from well-known
limitations: poor scalability on massively parallel
architectures and inability to handle non-periodic
systems efficiently. Real-space discretization methods
including those based on finite differences~\cite{Chelikowsky1994,Kronik2006},
wavelets~\cite{Genovese2008}, and finite
elements~\cite{Tsuchida1996,Pask2005,Motamarri2013}\cb{,}\cn{}
have emerged as scalable alternatives that accommodate
generic boundary conditions and are amenable to adaptive
resolution.

Among these, the finite-element (FE) basis, built from
compactly supported, piecewise-polynomial functions
constructed on an adaptively refined mesh, offers a
particularly compelling combination of properties for
large-scale DFT on modern heterogeneous architectures.
The locality of FE basis functions gives rise to sparse
discretized operators whose action can be evaluated locally at FE cell level, enabling fine-grained parallelism~\cite{Davydov2020,Fischer2020,Panigrahi2024} and
efficient GPU utilization through dense batched
matrix-matrix products~\cite{Das2022}.
The open-source DFT-FE code~\cite{Motamarri2020,Das2022}
exploits these features and has demonstrated exceptional
scalability for collinear-spin DFT, handling systems
with up to 100,000 electrons on 200,000 CPU cores and
40,000 GPUs~\cite{Das2019,Das2023}.

\subsection{Computational Challenges at Scale}
Extending DFT to incorporate noncollinear magnetism and
SOC introduces a qualitatively different computational
regime. The use of two-component complex spinors doubles
the dimension of the FE-discretized Hamiltonian from
$M$ to $2M$ (where $M$ is the number of FE grid points), and
the minimum number of eigenpairs required increases from
$N_{\mathrm{el}}/2$ (collinear) to $N_{\mathrm{el}}$ (noncollinear) all in
complex arithmetic. Complex floating-point operations
carry a higher cost ($\approx$ 3$\times$ to 4$\times$) than their real counterparts, and
together these factors increase the computational
prefactor by approximately $12\times$--$24\times$ relative to
standard collinear DFT calculations depending on the sizes of the material systems studied. Except for a few recent works~\cite{He2021,KodaliNC2025}, typical NC-SOC-DFT calculations with
state-of-the-art DFT codes are confined to systems of
$\sim$500--2000 electrons. Ref.~\cite{He2021} demonstrated large-supercell NC-SOC calculations with the plane-wave/PAW code VASP but did not introduce new eigensolver algorithms or report scalability, whereas Ref.~\cite{KodaliNC2025} established the finite-element NC-SOC framework up to 20{,}640 electrons on 512 GPUs as a proof of concept. Building on the latter, the present work adds matrix-free Poisson solves, mixed-precision and compressed-communication eigensolver kernels, and band-partitioned data layouts, extending NC-SOC-DFT to 78{,}988 electrons on 9{,}600 GPUs. Reaching the
$\mathcal{O}(10^4)$--$\mathcal{O}(10^5)$ electron
regime in a scalable manner is computationally very challenging. Prior DFT calculations reaching
scales of $\mathcal{O}(10^5)$ electrons were limited
to collinear-spin formulations~\cite{Gygi2006, Alvarez2008, Hasegawa2011, Ziogas2019, Fattebert2016, Das2019,Das2023}. Addressing this computational challenge demands co-designed
algorithmic and HPC innovations targeting two key components of the SCF
algorithm: the linear solver for DFT electrostatics (Equation~\eqref{eq:poisson_el})
and the eigensolver for the Kohn-Sham DFT
eigenproblem (Equation~\eqref{eq:ks_eigenproblem}). The eigensolver dominates the SCF
cost at the system sizes reported here, while the electrostatic solve accounts for a small
fraction of that cost for the largest systems but determines time-to-solution for smaller
systems and limits strong scaling at extreme node counts. Developing these innovations is
the focus of the current work.

\section{Innovations realized}
The NC-SOC-DFT equations described in Section~\rom{2} are solved by combining a higher-order adaptive finite-element (FE) discretization with efficient solver strategies and GPU-centric HPC innovations. We first describe the FE discretization and its challenges specific to NC-SOC, followed by the solver strategies and HPC innovations that enable large-scale calculations.

\subsection{Finite-element discretization for NC-SOC-DFT}
The two-component complex spinors are expanded in the FE basis as
\begin{equation}
  \Psi_{n}(\br) = \sum_{I=0}^{M-1}
  \begin{bmatrix} \psi^{I,\uparrow}_{n} \\ \psi^{I,\downarrow}_{n} \end{bmatrix}
  N_I(\br),
  \label{eqn:FEWfns}
\end{equation}
where $N_I(\br)$ are 3D tensor-structured FE polynomial basis functions~\cite{Hughes2000} constructed from 1D Lagrange polynomials of degree $p$ on Gauss--Lobatto--Legendre (GLL) nodes~\cite{Motamarri2013}, and $M$ denotes the number of FE nodes (grid points). This discretization transforms the DFT eigenproblem in \cref{eq:ks_eigenproblem} into a $2M\times 2M$ sparse generalized Hermitian eigenvalue problem, $\bH\bX = \bM\bX\bLam$, where $\bX=[\bx_1,\ldots,\bx_{N_e}]$ contains the FE coefficients of the spinor wavefunctions, while $\bLam=\operatorname{diag}(\epsilon_1,\ldots,\epsilon_{N_e})$, and the Hamiltonian $\bH = \bH^{\mathrm{loc}} + \bH^{\mathrm{nloc}}$ is a complex Hermitian matrix. The overlap matrix $\bM$, corresponding to the FE mass matrix, is real symmetric and positive definite. The local part $\bH^{\mathrm{loc}}$ includes the kinetic energy, effective potential, and exchange-correlation magnetic field ($\bB_{\mathrm{xc}}$) contributions, while $\bH^{\mathrm{nloc}}$ encodes the spin-orbit coupling via the separable pseudopotential. The overlap matrix $\bM$ has the form $\bM_{2I+\alpha,2J+\beta} = \delta_{\alpha\beta}\int N_I(\br)N_J(\br)\,d\br$, where $\alpha,\beta\in\{0,1\}$ index the two spin components. It is diagonal in spin indices and inherits the usual sparse structure of the FE mass matrix. All integrals are evaluated as sums over non-overlapping hexahedral FE cells $\Omega^{(e)}$ using Gauss--Legendre quadrature rules of sufficiently high order~\cite{KodaliNC2025}. Similarly, the total electrostatic potential $V_{\mathrm{el}}$ is obtained by solving the FE-discretized Poisson equation $\mathbf{K}\mathbf{V}_{\mathrm{el}} = \mathbf{f}$, where $\mathbf{K}$ is the discrete Laplacian operator and $\mathbf{f}$ is the assembled right-hand side of \cref{eq:poisson_el}.

\paragraph{Challenges specific to NC-SOC}
The NC-SOC discretization introduces several challenges beyond the collinear case. First, as stated before, the two-component spinor structure doubles the problem dimension from $M$ to $2M$ and requires complex arithmetic throughout, increasing the computational cost by approximately $12\times$--$24\times$. Second, evaluation of the exchange-correlation magnetic field $\bB_{\mathrm{xc}}$ for generalized gradient approximation (GGA) functionals requires gradients of the magnetization unit vector $\hat{\mathbf{m}} = \mathbf{m}/|\mathbf{m}|$, which become singular when $|\mathbf{m}| \to 0$. The standard divergence-theorem approach used in collinear calculations~\cite{Motamarri2020,Das2022} to recast Laplacians as gradients fails in the NC case due to this singularity. We address this by extending the White--Bird formalism~\cite{KodaliNC2025} to the FE setting, computing $V_{\mathrm{xc}}$ and $\bB_{\mathrm{xc}}$ directly at quadrature points from the discretized exchange-correlation energy, thereby avoiding the ill-defined gradient of $\hat{\mathbf{m}}$.  Third, the non-local pseudopotential operator acquires a full $2\times2$ spin-block structure that couples the two spinor components through the SOC channels, thereby requiring modified projection and assembly routines. The resulting discretized problem is solved via a self-consistent field (SCF) iteration using Anderson mixing with Resta-like preconditioning~\cite{Kim_2022}, where each SCF step requires: (i) solving the Poisson equation for $V_{\mathrm{el}}$, (ii) evaluating $V_{\mathrm{xc}}$ and $\bB_{\mathrm{xc}}$, and (iii) solving the $2M\times 2M$ generalized eigenproblem for $N_e$ eigenpairs. Step (iii) dominates the SCF cost at the system sizes considered in this work, while step (i) accounts for a smaller share of the largest calculations but limits strong scaling at extreme node counts. Both are described subsequently.
\subsection{Efficient and scalable solver strategies}

\subsubsection{Linear solver for electrostatic problem}
The FE-discretized Poisson problems in \cref{eq:poisson_el}
are solved with a preconditioned conjugate gradient (PCG) method. While the electrostatic solve constitutes a relatively small fraction of the SCF time for the largest systems considered in this work, it can be computationally expensive for smaller systems and a bottleneck to strong scaling at extreme node counts. PCG is inherently communication-intensive because each iteration combines global vector reductions with distributed applications of the discrete Laplace operator. In fact, each iteration requires two global \texttt{MPI\_Allreduce} operations, one for each of the inner products that set the step length and the search direction, together with nearest-neighbour ghost exchanges, and several thousand iterations may be required to reach the prescribed solver tolerance. As the number of MPI tasks increases and the degrees of freedom per GPU decrease, communication and synchronization increasingly dominate the PCG solve, causing it to lose parallel efficiency faster than the other components of SCF iteration, limiting the overall strong-scaling efficiency. To this end, the GPU-centric matrix-free kernels and fused CG iteration developed in this work reduce the per-iteration cost and data movement of the Poisson solve, improving time-to-solution over state-of-the-art PCG implementations and thereby enhancing the scalability of the overall SCF iteration.

A dominant computational cost in each PCG iteration is the application of the discrete Laplace operator $\mathbf{K}$ to
a trial vector. We implement this matrix-vector product as a matrix-free GPU kernel that exploits the tensor-product structure of the FE basis. Since $d$-dimensional basis functions factor into products of 1D shape functions, the stiffness matrix action decomposes into a sequence of 1D contractions, reducing the per-element cost from $\mathcal{O}(p^{2d})$ to
$\mathcal{O}(dp^{d+1})$ for a finite-element polynomial order $p$. The entire sequence of extraction, contractions, Jacobian application, and assembly is fused into a single GPU kernel, hereafter the $\mathbf{K}\mathbf{x}$ kernel,
with the kernel-level design detailed in \cref{sec:mfcg}.

For preconditioning, we employ a low-degree Chebyshev polynomial approximation to the inverse
of the discrete Laplace operator. Though matrix-free $h$- and $p$-multigrid preconditioners are attractive alternatives, with iteration counts largely independent of mesh size and polynomial order~\cite{KronbichlerWall2018,Fischer2020}, device-resident implementations have so far been demonstrated at the node level~\cite{KronbichlerLjungkvist2019,Cui2025}, while the corresponding large-scale distributed results are reported on CPUs~\cite{Davydov2020,Munch2023}, and adapting these methods to multinode GPU architectures remains an active research direction~\cite{Wichrowski2025}. A degree-$m$ Chebyshev preconditioner reuses the fused $\mathbf{K}\mathbf{x}$ kernel and requires $m-1$ additional ghost exchanges per PCG iteration, but no global reductions. The low-degree polynomial for the preconditioner is therefore well-suited to the latency-bound regime considered here, improving the convergence rate while avoiding the global reductions associated with coarse-grid solves in a multigrid preconditioner. The Chebyshev recurrence in the preconditioner action requires repeated applications of $\mathbf{K}$ and diagonal scaling, both of which are performed using the same matrix-free infrastructure. All PCG operator applications, preconditioner operations, residual evaluations, and communication use FP64 in the current work.

\subsubsection{Chebyshev filtered subspace iteration for generalized eigenproblems}

The FE-discretized KS eigenproblem in \cref{eq:ks_eigenproblem} yields a sparse generalized Hermitian eigenvalue problem
\begin{equation}
  \bH\bx_j = \epsilon_j\bM\bx_j, \quad j=1,\dots,N_e,
  \label{eqn:ghep}
\end{equation}
where $\bH\in\mathbb{C}^{2M\times 2M}$ is the discretized Hamiltonian, $\bM\in\mathbb{R}^{2M\times 2M}$ is the symmetric positive-definite FE overlap (mass) matrix, and the factor of two in the matrix dimensions accounts for the two spinor components. In previous collinear DFT-FE implementations~\cite{Das2023,Motamarri2020}, $\bM$ was rendered diagonal via a GLL-integrated or lumped mass approximation, thereby reducing \cref{eqn:ghep} to a standard eigenproblem. For NC-SOC, we retain the full mass matrix in the generalized eigenproblem to avoid introducing additional approximation error into the magnetization density~\cite{KodaliNC2025}.

Our eigensolver is based on Chebyshev filtered subspace iteration (ChFSI)~\cite{Saad2006a}. At eigensolver iteration $i$, a degree-$q$ Chebyshev filter is applied to the current Ritz vectors $\bX^{(i)}$, with associated Ritz values $\bLam^{(i)}$, to construct the filtered subspace $\bY_q^{(i)}=C_q(\bM^{-1}\bH)\bX^{(i)}$, followed by a Rayleigh-Ritz (RR) step that produces the updated Ritz vectors $\bX^{(i+1)}$ and Ritz values $\bLam^{(i+1)}$. The filtered subspace is constructed via the three-term recurrence:
\begin{align}
  \bY_{k+1}^{(i)} = \frac{2\sigma_{k+1}}{e}\bM^{-1}\bH\,\bY_k^{(i)}
    - \frac{2\sigma_{k+1}c}{e}\bY_k^{(i)}
    - \sigma_k\sigma_{k+1}\bY_{k-1}^{(i)},
  \label{eqn:ChFSIrecurrence}
\end{align}
where $k=1,\ldots,q-1$, $c=(\lambda_{\max}+\lambda_T)/2$, $e=(\lambda_{\max}-\lambda_T)/2$, and the scaling coefficients satisfy $\sigma_1=e/(\lambda_{\min}-c)$ and $\sigma_{k+1}=(2/\sigma_1-\sigma_k)^{-1}$. The filter parameters $\lambda_{\min}$ and $\lambda_{\max}$ are lower and upper bounds, respectively, on the spectrum of $\bM^{-1}\bH$, and $\lambda_T$ estimates the first unwanted eigenvalue. Here $T_k$ is the degree-$k$ Chebyshev polynomial of the first kind, and $C_k(x)=T_k((x-c)/e)/T_k((\lambda_{\min}-c)/e)$ uses the affine mapping of $[\lambda_T,\lambda_{\max}]$ to $[-1,1]$ and is normalized at $\lambda_{\min}$. The recurrence starts with $\bY_0^{(i)}=\bX^{(i)}$ and $\bY_1^{(i)}=\sigma_1(\bM^{-1}\bH-c\mathbf{I})\bX^{(i)}/e$. Each recurrence step requires a Hamiltonian--multivector product followed by the action of $\bM^{-1}$. Since applying $\bM^{-1}$ exactly is prohibitively expensive at scale, we use $\bD^{-1}\approx\bM^{-1}$ during filtering, where $\bD$ is either a GLL-integrated or lumped mass matrix. If $\bY_q^{(i)}$ and $\underline{\bY}_q^{(i)}$ denote the filtered subspaces obtained using the exact and approximate actions, respectively, the resulting error $\bDelta_q^{(i)} = \underline{\bY}_q^{(i)} - \bY_q^{(i)}$ has a bound containing a term independent of the eigenproblem residual~\cite{rchfsiCPC}, and ChFSI can therefore stagnate at a residual floor determined by the approximation quality.

\paragraph{Residual-based reformulation (R-ChFSI)}
To overcome this limitation, we employ the residual-based ChFSI (R-ChFSI) method~\cite{rchfsiCPC}, which reformulates the Chebyshev recurrence in terms of weighted residuals $\bZ_k^{(i)} = \bD(\bY_k^{(i)} - \bX^{(i)}\bLam_k^{(i)})$, where $\bLam_k^{(i)}=C_k(\bLam^{(i)})$ is the Chebyshev-filtered Ritz-value matrix. Here $\bR^{(i)}=\bH\bX^{(i)}-\bM\bX^{(i)}\bLam^{(i)}$ denotes the eigenproblem residual. As the eigenpairs converge, these weighted residuals approach zero, and the filtered-subspace error resulting from reduced-precision Hamiltonian--multivector products or from using $\bD^{-1}$ in place of $\bM^{-1}$ has a bound proportional to the residual norm $\|\bR^{(i)}\|$. Using $\bD\bM^{-1}\approx\mathbf{I}$ gives the following practical R-ChFSI recurrence in place of \cref{eqn:ChFSIrecurrence}:
\begin{multline}
  \bZ_{k+1}^{(i)} = \frac{2\sigma_{k+1}}{e}\bH\bD^{-1}\bZ_k^{(i)}
    - \frac{2\sigma_{k+1}c}{e}\bZ_k^{(i)} \\
    - \sigma_k\sigma_{k+1}\bZ_{k-1}^{(i)}
    + \frac{2\sigma_{k+1}}{e}\bR^{(i)}\bLam_k^{(i)},
  \label{eqn:rChFSIrecurrence}
\end{multline}
where $k=1,\ldots,q-1$. We note that $\bR^{(i)}$ is computed once per eigensolver iteration in double precision, and $\bLam_k^{(i)}$ is propagated in double precision using the same three-term Chebyshev recurrence. The recurrence starts with $\bZ_0^{(i)}=\bzero$ and $\bZ_1^{(i)}=\sigma_1\bR^{(i)}/e$. After $q$ steps, the filtered vectors are recovered as $\bY_q^{(i)} = \bD^{-1}\bZ_q^{(i)} + \bX^{(i)}\bLam_q^{(i)}$. This reformulation is critical for the NC-SOC eigenproblem: it permits the use of a simple diagonal $\bD^{-1}$ in place of the exact $\bM^{-1}$ while retaining convergence to the eigenpairs of the generalized eigenproblem with the full mass matrix. The overall R-ChFSI eigensolver procedure is summarized in \cref{alg:RChFSI}.

\begin{algorithm}[!htbp]
\caption{R-ChFSI eigensolver for the generalized Hermitian eigenproblem}
\label{alg:RChFSI}
\begin{algorithmic}[1]
\Require $\bH\in\mathbb{C}^{2M\times 2M}$, $\bM\in\mathbb{R}^{2M\times 2M}$; initial Ritz pairs $(\bX^{(0)},\bLam^{(0)})$; number of eigenpairs $N_e$; polynomial degree $q$; initial filter-parameter estimates $\lambda_{\min},\lambda_{\max},\lambda_T$; diagonal approximation $\bD^{-1}\approx\bM^{-1}$; tolerance $\tau$
\Ensure Approximate Ritz pairs $(\bX,\bLam)$
\State Set $i\gets 0$
\While{$\max_j\|\bH\bx_j^{(i)}-\epsilon_j^{(i)}\bM\bx_j^{(i)}\| \geq \tau$}
  \State Compute residual $\bR^{(i)}=\bH\bX^{(i)}-\bM\bX^{(i)}\bLam^{(i)}$ \hfill\Comment{FP64}
  \State \textbf{R-ChFSI filter:} Compute $\bZ_q^{(i)}$ via \cref{eqn:rChFSIrecurrence} \hfill\Comment{low-precision Hamiltonian--multivector products (FP32 or TF32)}
  \State Recover $\bY_q^{(i)}=\bD^{-1}\bZ_q^{(i)}+\bX^{(i)}\bLam_q^{(i)}$
  \State \textbf{Rayleigh-Ritz (RR) step}:\\
  \hspace{0.25in} Solve subspace eigenproblem for $\bQ$ and $\bLam^{(i+1)}$:\\ \hspace{0.25in} ${\bY_q^{(i)}}^\dagger\bH\bY_q^{(i)}\bQ={\bY_q^{(i)}}^\dagger\bM\bY_q^{(i)}\bQ\bLam^{(i+1)}$\\
  \hspace{0.25in} Subspace rotation (SR): $\bX^{(i+1)}\gets\bY_q^{(i)}\bQ$
  \State Update $\lambda_{\min}$ and $\lambda_T$ from $\bLam^{(i+1)}$
  \State $i\gets i+1$
\EndWhile
\end{algorithmic}
\end{algorithm}

R-ChFSI enables two filtering optimizations used in this work. First, the Hamiltonian--multivector product in \cref{eqn:rChFSIrecurrence} can be evaluated in single or lower precision (FP32 or TF32), since the residual-proportional error bound~\cite{rchfsiCPC} ensures that the reduced-precision errors do not impede convergence. Second, $\bZ_k^{(i)}$, communicated during the distributed Hamiltonian--multivector product, can be compressed using block floating-point (BFP) at compression ratios of 4$\times$--6.4$\times$ relative to FP64, further reducing data movement costs. Both strategies are detailed in \cref{sec:rchfsi-lowprec,sec:bfp-compression}.

\subsection{HPC Innovations}

\subsubsection{GPU-accelerated matrix-free conjugate gradient solver}
\label{sec:mfcg}

Our Preconditioned Conjugate Gradient (PCG) solver implementation keeps all vector operations, inner products and matrix-vector products on
the GPU for the entire solve, so no vector is copied back to the host between
iterations. The iteration itself is standard PCG with its usual two synchronization
points. Fusing the inner products $\mathbf{r}\cdot\mathbf{z}$ and
$\mathbf{r}\cdot\mathbf{r}$ into one device kernel lets the residual norm used for the
convergence test be carried by the second \texttt{MPI\_Allreduce} rather than requiring a
third, and the updates to $\mathbf{x}$ and $\mathbf{r}$ are fused into a single pass over
the vectors (\cref{alg:pcg_fusion}).

\definecolor{CGglobal}{RGB}{196,78,82}       %
\definecolor{CGregel}{RGB}{58,110,190}        %
\definecolor{CGshared}{RGB}{214,140,35}       %
\definecolor{CGconst}{RGB}{55,155,90}         %
\definecolor{CGmpi}{RGB}{130,65,175}          %
\definecolor{CGkernel}{RGB}{30,145,150}       %
\definecolor{CGprec}{RGB}{175,80,140}         %
\definecolor{CGsync}{RGB}{230,185,60}         %

\begin{algorithm}[!h]
\caption{GPU-Accelerated Preconditioned CG Iteration (Fused Kernels)}%
\label{alg:pcg_fusion}
\begin{algorithmic}[1]
\Require $\mathbf{K}$, $\mathbf{P}^{-1}$, tolerance $\varepsilon$; $\mathbf{x}$, $\mathbf{r}=\mathbf{K}\mathbf{x}-\mathbf{f}$, $\mathbf{p}$, $\delta=\mathbf{r}\cdot\mathbf{z}$ from the previous iteration
\Ensure Updated $\mathbf{x}$, $\mathbf{r}$, $\mathbf{p}$, $\delta_{\text{new}}$
\State \textbf{$\mathbf{K}\mathbf{x}$ kernel:} $\mathbf{w} \leftarrow \mathbf{K}\mathbf{p}$
\State \textbf{Dot kernel:} $\eta \leftarrow \mathbf{p}\cdot\mathbf{w}$
\State \texttt{MPI\_Allreduce} ($\eta$, 1 scalar);\quad $\alpha \leftarrow \delta / \eta$
\State \textbf{Fused update kernel:} $\mathbf{x} \mathrel{+}= \alpha\mathbf{p}$,\; $\mathbf{r} \mathrel{+}= \alpha\mathbf{w}$
\State \textbf{Chebyshev precond.:} $\mathbf{z} \leftarrow \mathbf{P}^{-1}\mathbf{r}$
\State \textbf{Fused dot kernel:} $(\delta_{\text{new}},\, \gamma) \leftarrow (\mathbf{r}\cdot\mathbf{z},\; \|\mathbf{r}\|^2)$
\State \texttt{MPI\_Allreduce} ($[\delta_{\text{new}},\gamma]$, 2 scalars)
\State $\beta \leftarrow \delta_{\text{new}} / \delta$;\quad $\delta \leftarrow \delta_{\text{new}}$
\State $\mathbf{p} \leftarrow -\mathbf{z} + \beta\,\mathbf{p}$
\State \textbf{if} $\sqrt{\gamma} < \varepsilon$ \textbf{then} converged \textbf{end if}
\end{algorithmic}
\end{algorithm}

\begin{figure}[!ht]
\centering
\includegraphics[width=0.48\textwidth]{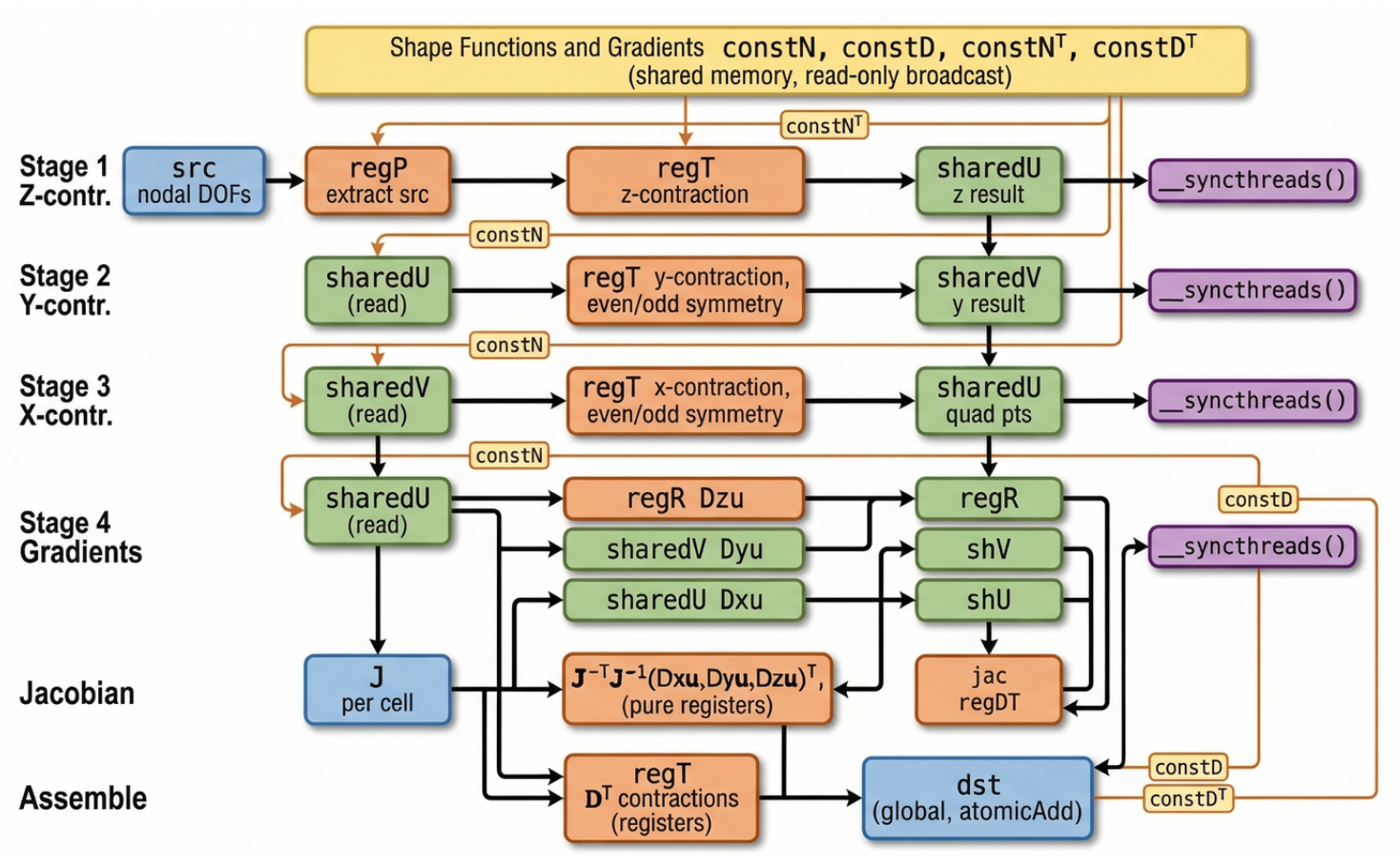}
\caption{Hybrid register--shared memory strategy for the matrix-free $\mathbf{K}\mathbf{x}$ kernel; intermediate values remain in registers unless needed across threads.}%
\label{fig:hybrid_reg_share}
\end{figure}

\paragraph{The $\mathbf{K}\mathbf{x}$ Kernel: Fusing Extraction, Tensor Contractions, and Assembly}
The dominant cost is the matrix-free action of $\mathbf{K}$, which applies the
finite-element stiffness matrix without ever forming it~\cite{Panigrahi2024}. A single kernel
launch first extracts the FE cell-level nodal values from the global FE degree-of-freedom
vector (an index-gather through the local-to-global map), then performs six batched
tensor contractions covering interpolation of nodal values to quadrature points,
application of the per-cell Jacobian, and integration back to the nodal basis, and
finally assembles the cell contributions back into the global output vector. Three levels of fusion keep the intermediates out of global memory
(see Fig.~\ref{fig:hybrid_reg_share}).

\begin{itemize}

\item \textit{Fusing extraction with the first contraction:} Each thread gathers the
nodal values along one $z$-line of the cell straight from the global vector into a
private register array and applies the $z$-contraction as the values arrive, so the
extraction is never written out as a separate cell-local buffer and memory latency is
overlapped with arithmetic. This pays off most at higher polynomial orders, where the kernel is
memory-bound. After staging through shared memory, the $y$- and
$x$-contractions apply the small dense 1D FE shape-function matrices along one direction of
the cell at a time. These matrices hold the shape function values and their gradients at
the quadrature points. The GLL nodes and quadrature points are symmetric about the cell midpoint, so reversing
both index ranges leaves the value matrix unchanged and flips the sign of the gradient
matrix. An even/odd decomposition exploits this symmetry to halve the multiplication
count~\cite{Solomonoff1992,Kopriva2009}.

\item \textit{Hybrid register--shared memory strategy:} Values that require no
inter-thread communication stay in registers, including the gradient contractions and
the Jacobian application, lowering shared memory pressure and raising GPU occupancy.

\item \textit{Fusing the last contraction with assembly:} Register-resident outputs are
accumulated into the output array by atomic additions, eliminating one global memory
pass.
\end{itemize}

\paragraph{Preconditioner Application and Inner Products}
The preconditioner applies a degree-$m$ Chebyshev polynomial to the diagonally scaled
Laplacian, $\mathbf{P}^{-1} = \chi(\mathbf{K}_{\mathrm{diag}}^{-1}\mathbf{K})\,
\mathbf{K}_{\mathrm{diag}}^{-1}$ with
$\mathbf{K}_{\mathrm{diag}}=\mathrm{diag}(\mathbf{K})$, where $\chi$ approximates
$\lambda^{-1}$ over the spectrum of $\mathbf{K}_{\mathrm{diag}}^{-1}\mathbf{K}$, so that
$\mathbf{P}^{-1}\approx\mathbf{K}^{-1}$. The spectral bounds are estimated once per solve
by a short Lanczos run on the scaled operator. Each Chebyshev step
applies $\mathbf{K}$ once and fuses the residual computation, the diagonal scaling, the
direction rescaling, and accumulation of the output into one register-resident kernel. The
inner products required in CG iteration use a five-phase tree reduction: register accumulators per thread, a
warp-level reduction through shuffle intrinsics, a shared memory tree across warps, a
block-level partial sum, and two atomic additions per threadblock into pre-zeroed global
accumulators. Almost all of the arithmetic thus stays in registers and first-level
caches.

\subsubsection{Reduced precision arithmetic for matrix--multivector products in R-ChFSI}
\label{sec:rchfsi-lowprec}
For medium-scale systems, the Hamiltonian--multivector product is the dominant cost of each R-ChFSI iteration. A direct sparse-matrix times dense-matrix (spMM) kernel is memory-bandwidth bound and achieves poor utilization on modern GPUs. We instead exploit the compact support of the FE basis to recast the global spMM into batched cell-level dense matrix--matrix products~\cite{KodaliNC2025,Panigrahi2024} that are compute-bound and map directly onto high-throughput \texttt{GEMM} hardware units. Each FE cell $\Omega^{(e)}$ carries a precomputed dense local Hamiltonian ${\bH^{\mathrm{loc}}}^{(e)}$ of size $2n_p^3 \times 2n_p^3$, where $n_p = p+1$ is the number of finite-element nodes per direction and the factor of 2 accounts for the spinor components. In the NC-SOC case, the $2\times2$ spin-block structure is
\begin{equation}
  {\bH^{\mathrm{loc}}}^{(e)} =
  \begin{bmatrix}
    \bH_{\mathrm{scal}}^{(e)} + \bH_{B_z}^{(e)} & \bH_{B_x}^{(e)} - i\bH_{B_y}^{(e)} \\
    \bH_{B_x}^{(e)} + i\bH_{B_y}^{(e)} & \bH_{\mathrm{scal}}^{(e)} - \bH_{B_z}^{(e)}
  \end{bmatrix},
  \label{eqn:cellHam}
\end{equation}
where $\bH_{\mathrm{scal}}^{(e)}$ encodes the kinetic and effective potential contributions and $\bH_{B_d}^{(e)}$ ($d=x,y,z$) encode the exchange-correlation magnetic field components. The product $\bY = \bH\bX$ is evaluated in four stages: (i)~cell-level multivector extraction $\bX^{(e)} = \mathcal{R}^{(e)}\bX$ via index-gather into contiguous GPU buffers, where $\mathcal{R}^{(e)}$ restricts a global multivector to cell $e$; (ii)~batched strided \texttt{GEMM} over all local cells, $\bY^{(e)} = {\bH^{\mathrm{loc}}}^{(e)}\bX^{(e)}$, where each batch element is a $(2n_p^3)\times(2n_p^3) \;\times\; (2n_p^3)\times N_b$ product with $N_b$ wavefunctions; (iii)~assembly via index-scatter with atomic additions $\bY \mathrel{+}= {\mathcal{R}^{(e)}}^T\bY^{(e)}$; and (iv)~non-local pseudopotential contributions are similarly evaluated using batched \texttt{GEMM}s with cell-level projector matrices ${\bP^a}^{(e)}$ and the SOC coefficient matrices $\bDelta^{\gamma_a}$ (where $\gamma_a$ denotes the type of atom $a$). The extraction--GEMM--assembly pipeline is overlapped with nearest-neighbour MPI ghost exchange (scatter/gather of ghost values) for the domain-decomposed FE mesh, hiding communication latency behind the compute-intensive cell \texttt{GEMM}s.

This cell-matrix formulation directly enables mixed-precision computation in the R-ChFSI filtering step. After the FP64 assembly of ${\bH^{\mathrm{loc}}}^{(e)}$, FP32 copies are generated via a single device-to-device cast at negligible cost. During R-ChFSI filtering, all batched \texttt{GEMM}s, which constitute the bulk of the FLOP count, use FP32 operands. On Intel PVC GPUs, the \texttt{GEMM} calls use TF32 tensor operations via the \texttt{oneMKL} interface, whereas on Frontier's AMD MI250X GPUs the \texttt{hipBLAS} interface executes them in FP32 because TF32 is not supported. The extraction, scatter/gather, and assembly kernels similarly operate on FP32 multivectors, halving global memory traffic relative to FP64. Only the eigenproblem residual $\bR^{(i)}$ and the eigenvalue recurrence $\bLam_k^{(i)}$ are retained in FP64, ensuring that the R-ChFSI convergence bound holds and the overall solver delivers double-precision accuracy.

\subsubsection{Block floating-point compression for point-to-point (P2P) communication}
\label{sec:bfp-compression}

\begin{figure}[!h]
    \centering
    \includegraphics[width=0.45\textwidth]{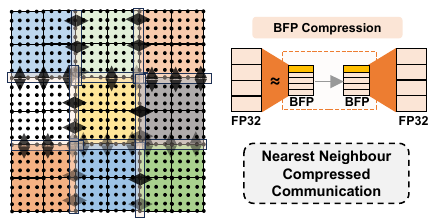}
    \caption{BFP fixed-rate compressed communication across domain boundaries with data compressed/decompressed on GPUs; colors denote MPI ranks.}
    \label{fig:P2P}
\end{figure}

In a finite element (FE) discretization, the basis functions are locally supported, so point-to-point (P2P) communication is confined to nearest neighbours across domain boundaries. Each application of the Hamiltonian to trial residual vectors in the Chebyshev recurrence in Equation~\eqref{eqn:rChFSIrecurrence} requires two ghost exchanges: a ghost update, in which each rank sends its owned boundary values to neighbouring ranks holding them as ghosts, and a ghost accumulation, in which the accumulated ghost contributions are returned to the owning rank (Fig.~\ref{fig:P2P}). For two ranks sharing $n_\Gamma$ boundary nodes, each exchange communicates $f \times n_\Gamma \times N_b$ values, with $N_b$ denoting the wavefunction block size and $f=1$ for real and $f=2$ for complex wavefunctions. With $n_{\mathrm{bits}}$ bits used per value, each exchange therefore communicates $\left\lceil (f n_\Gamma N_b n_{\mathrm{bits}})/8 \right\rceil$ bytes, making the communication volume directly proportional to $n_{\mathrm{bits}}$ (e.g., $n_{\mathrm{bits}}=32$ for FP32).

Residual-based Chebyshev filtering, which is tolerant to inexact matrix--multivector products, enables the use of FP32 precision for both computation and communication, accelerating the filtering phase by up to $2\times$ compared to FP64 while maintaining similar convergence. Furthermore, using BF16 precision for P2P communication alone provides additional performance gains of $1.3\times$--$1.5\times$, depending on the scaling regime and the underlying hardware \cite{rchfsiCPC, ramakrishnan2026acceleratingfiniteelementbasedprojectoraugmentedwave}. However, for material systems that require harder pseudopotentials, the limited mantissa precision of BF16 can increase the number of SCF iterations, offsetting any gains from reduced-precision communication. Although FP16 provides greater precision than BF16 with the same 16-bit representation, its narrower dynamic range makes it more susceptible to overflow. These limitations motivate the use of numerical formats that require 16 or fewer bits per value while providing higher accuracy than standard 16-bit datatypes without compromising SCF convergence. This is particularly important in communication-bound regimes. For NC-SOC-DFT systems, the computationally intensive $\bH\bX$ operation can be effectively accelerated on TF32 tensor cores. However, the resulting increase in computational throughput shifts the GPUs toward a communication-bound regime, degrading strong scaling and motivating the use of lower-precision formats to reduce communication volume.

\begin{figure}[!h]
    \centering
    \includegraphics[width=0.48\textwidth]{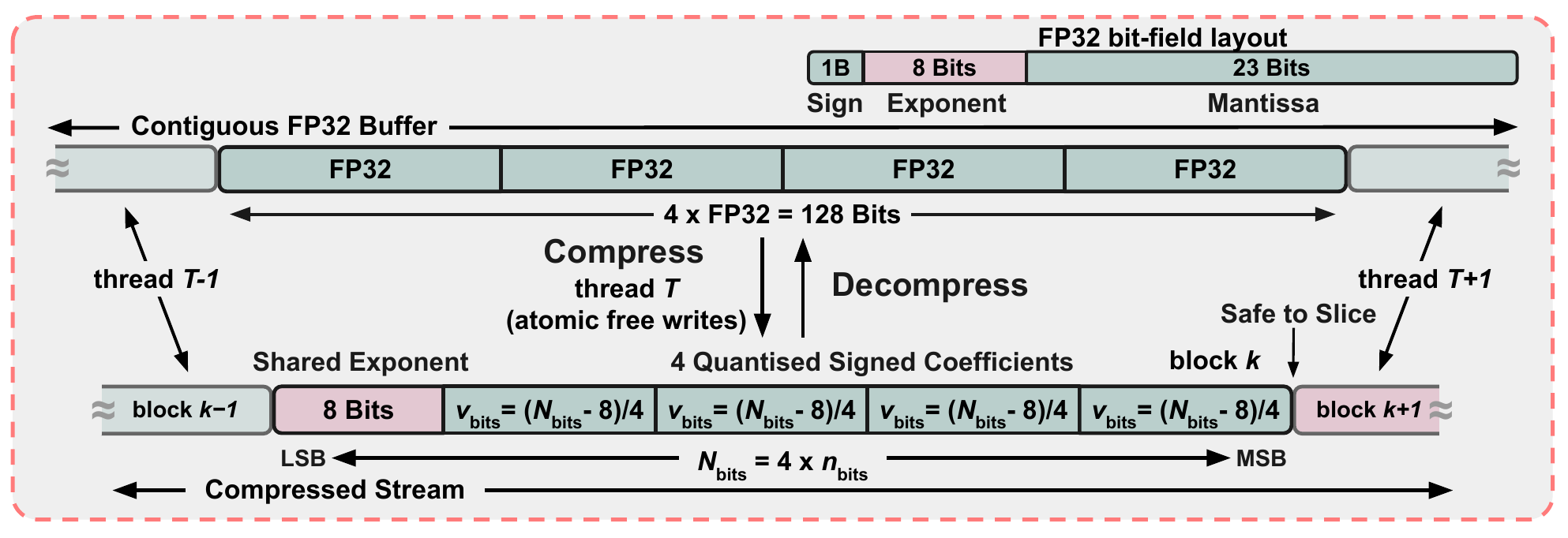}
\caption{BFP compressed-stream layout: each thread encodes/decodes 4 FP32 values into a fixed-size block; no storage word is straddled by two blocks.}
    \label{fig:layout}
\end{figure}

To this end, we propose using the block floating-point (BFP) representation to
compress exchanged values at a predetermined fixed rate with negligible
encoding and decoding overheads. BFP formats have long been used in
signal processing systems, and in recent years, BFP and closely related
formats such as Microscaling (MX) have gained increasing popularity in AI
workloads, particularly for LLMs~\cite{3326943.3326985,
rouhani2023microscalingdataformatsdeep}. BFP uses a single shared
exponent (determined by the largest absolute value in a block) for a group of
values, with each value then quantized using this shared exponent and
represented by a signed integer of the same width. Our lightweight BFP implementation maps four FP32 values to $N_{\mathrm{bits}} = 4 \times n_{\mathrm{bits}}$ bits. Because the exponent is shared across the four values, $n_{\mathrm{bits}}$ is an amortized per-value rate, whereas $N_{\mathrm{bits}}$ is the actual size of the compressed four-value word, whose lower $8$ bits store a shared biased exponent spanning the full FP32 exponent range, while the remaining $N_{\mathrm{bits}}-8$ bits are divided equally among four signed coefficients, each of width $v_{\mathrm{bits}} = \frac{N_{\mathrm{bits}}-8}{4} = n_{\mathrm{bits}}-2$
(Fig.~\ref{fig:layout}). Thus, $n_{\mathrm{bits}} = 16$, $12$, and $10$
correspond to $v_{\mathrm{bits}} = 14$, $10$, and $8$, respectively. At $n_{\mathrm{bits}} = 16$, each value uses $v_{\mathrm{bits}} = 14$ bits for its quantized signed coefficient, providing nearly twice as many significand bits as BF16 at the same $16$ bits per value. The shared exponent $e_{\max}$ satisfies $2^{e_{\max}-1} \leq \max_i |x_i| < 2^{e_{\max}}$. A block whose four values are all zero is flagged by a zero exponent field. Otherwise, the biased exponent $e=e_{\max}+127$ is stored, and each value is quantized as
$q_i = \operatorname{round}\!\left(x_i \times 2^{\,v_{\mathrm{bits}}-1-e_{\max}}\right)$, $q_i \in \bigl[-2^{v_{\mathrm{bits}}-1},\,2^{v_{\mathrm{bits}}-1}-1\bigr]$.
Round-to-nearest avoids the systematic bias of truncation. Only the upper bound requires clamping, since $|x_i|<2^{e_{\max}}$ bounds $q_i$ from below by construction. Each $q_i$ is stored using its low $v_{\mathrm{bits}}$ bits in two's complement, and the dropped high bits are copies of the sign bit, which are restored during decoding. Decoding recovers $e_{\max}=e-127$ and reconstructs $\tilde{x}_i = q_i \times 2^{e_{\max}-v_{\mathrm{bits}}+1}$.
Thus, encoding consists of exponent extraction, quantization, and bit packing while decoding only requires unpacking and rescaling. Any error due to finite precision is introduced during quantization at encode time while decoding exactly reconstructs the quantized representation. BFP is well suited for our application since exchanged residual values associated with neighbouring nodes typically have similar magnitudes and tend toward zero. Unlike ZFP~\cite{6876024}, which applies a decorrelating transform and embedded bit-plane coding to assign bits adaptively across coefficients to accommodate high-variability data, thereby incurring higher arithmetic cost and potential thread divergence on GPUs, our BFP implementation incurs negligible encoding and decoding overhead. Compression techniques are uncommon in scientific computing, particularly in electronic structure calculations, where stringent accuracy requirements limit their use, while their adoption here is enabled by the robustness of R-ChFSI.

Each GPU thread encodes or decodes a four-value block, providing fine-grained parallelism (see Fig.~\ref{fig:layout}). For the bit rates used here $n_{\mathrm{bits}} \in \{16, 12, 10\}$, $N_{\mathrm{bits}}$ corresponds to a whole number of native machine stores, so each thread writes its encoded block to a private region of the compressed byte stream: no storage word is straddled by two blocks, and encoding requires no atomics. At $n_{\mathrm{bits}} = 16$, a block is stored as a single \texttt{uint64\_t}, giving fully coalesced writes. At $n_{\mathrm{bits}} = 12$ and $10$, a block is stored using three \texttt{uint16\_t} or five \texttt{uint8\_t} stores, respectively, resulting in less coalesced but still conflict-free writes. In contrast, ZFP packs successive blocks back-to-back into a contiguous $64$-bit word stream, requiring atomic writes.

Choosing the wavefunction block size $N_b$ as a multiple of four ensures that no four-value block straddles a message boundary, so each rank sends and receives only complete blocks. Moreover, since $n_{\mathrm{bits}}$ is even for all supported rates, each block also occupies an exact integral number of bytes, guaranteeing byte alignment for every message. Hence, the message exchanged between two ranks is exactly $(n_\Gamma N_b n_{\mathrm{bits}})/8$ bytes ($\times 2$ for complex). Compressed data for each neighbour is stored contiguously without padding, with each neighbour's offset determined by the prefix sum of the preceding message sizes, allowing the compressed stream to be sliced safely.

Compression and decompression kernels are fused into the two ghost exchanges rather than performed as separate stages. During the ghost update, the send side gathers the entries for each neighbour and encodes them in the same pass, writing directly to the send buffer without forming an intermediate uncompressed copy, while the receiving rank decodes the incoming stream into its contiguous ghost region. During ghost accumulation, the ghost region is already contiguous and therefore requires no gather, so it is compressed directly into the send buffer, while the receiving rank decodes the data and atomically scatter-adds it to the corresponding owned entries. Thus, the fused GPU kernels process the data in a single pass, combining compression/decompression with data movement without additional memory allocation or kernel launches.

\subsubsection{Rayleigh-Ritz step} The computational bottleneck in Algorithm~\ref{alg:RChFSI} for large system sizes is the Rayleigh-Ritz (RR) procedure, whose computational cost scales cubically with system size. The RR procedure involves three critical steps: (i) \texttt{RR-Proj}: projecting the FE-discretized Hamiltonian ($\bH$) and overlap ($\bM$) matrices onto the Chebyshev filtered subspace, (ii) \texttt{RR-diag}: solving the subspace-projected generalized eigenvalue problem using ELPA~\cite{elpa}, (iii) \texttt{RR-SR}: a subspace rotation procedure to obtain approximate eigenvectors of the original eigenproblem. Among these steps, the \texttt{RR-Proj} and \texttt{RR-SR} steps are dominated by dense matrix-matrix multiplications and are implemented using Level-3 \texttt{BLAS} (e.g., \texttt{GEMM}) kernels. These operations are compute-intensive and exhibit high arithmetic intensity, making them well-suited for efficient execution on modern GPU architectures.

To improve scalability and reduce memory overhead, we employ a blockwise formulation~\cite{Das2022} for the RR step. In particular, the projection and subspace rotation operations are carried out in blocks of wavefunctions, thereby limiting peak memory usage and enabling better cache and device memory utilization. Furthermore, this blockwise strategy facilitates overlapping communication with computation by pipelining data movement with \texttt{GEMM} operations. In the \texttt{RR-SR} step, the subspace rotation $\bY_{q}\bQ$ is decomposed as
\[
\bY_q\bQ = \bY_q\bQ_{\mathrm{diag}} + \bY_q\bQ_{\mathrm{non\text{-}diag}}
\]
where $\bQ_{\mathrm{diag}}$ and $\bQ_{\mathrm{non\text{-}diag}}$ denote the diagonal and off-diagonal components of $\bQ$, respectively. A mixed-precision strategy is employed wherein $\bY_q\bQ_{\mathrm{non\text{-}diag}}$ is evaluated in \texttt{FP32} arithmetic, while $\bY_q\bQ_{\mathrm{diag}}$ is computed in \texttt{FP64} arithmetic. The above strategies reduce the computational prefactor associated with the cubic-scaling Rayleigh-Ritz procedure.

\subsubsection{Band-partitioning approach}
\label{sec:band_partitioning}

Under strong scaling (fixed problem size), domain decomposition across MPI tasks increases communication overhead during the Chebyshev filtering step, as the degrees of freedom per GPU (DoFs/GPU) decrease with increasing MPI task count. Techniques such as communication-computation overlap and compressed communication, employed in this work, alleviate this overhead but do not sufficiently mitigate the bottleneck in the extreme-scaling regime, where overall throughput remains constrained by inter-GPU communication latency and bandwidth. However, the reduction in DoFs/GPU accompanying strong scaling also increases the available memory per GPU. The band-partitioning strategy proposed here exploits this additional memory to reduce or eliminate communication across GPU groups during the filtering step.

\begin{figure}[!h]
    \centering
        \begin{tikzpicture}[scale=0.7, every node/.style={font=\scriptsize}]
            \draw[step=1cm,gray,thin] (0,0) grid (5,3);
            \fill[blue!20] (0,0) rectangle (1,3);
            \fill[green!20] (1,0) rectangle (2,3);
            \fill[yellow!20] (2,0) rectangle (3,3);
            \fill[orange!20] (3,0) rectangle (4,3);
            \fill[red!20] (4,0) rectangle (5,3);
            \draw[thick] (0,0) -- (0,3);
            \draw[thick] (1,0) -- (1,3);
            \draw[thick] (2,0) -- (2,3);
            \draw[thick] (3,0) -- (3,3);
            \draw[thick] (4,0) -- (4,3);
            \draw[thick] (5,0) -- (5,3);
            \draw[thick] (0,0) -- (5,0);
            \draw[thick] (0,1) -- (5,1);
            \draw[thick] (0,2) -- (5,2);
            \draw[thick] (0,3) -- (5,3);
            \node at (0.5, 2.5) {P(1,1)};
            \node at (1.5, 2.5) {P(1,2)};
            \node at (2.5, 2.5) {P(1,3)};
            \node at (3.5, 2.5) {P(1,4)};
            \node at (4.5, 2.5) {P(1,5)};
            \node at (0.5, 1.5) {P(2,1)};
            \node at (1.5, 1.5) {P(2,2)};
            \node at (2.5, 1.5) {P(2,3)};
            \node at (3.5, 1.5) {P(2,4)};
            \node at (4.5, 1.5) {P(2,5)};
            \node at (0.5, 0.5) {P(3,1)};
            \node at (1.5, 0.5) {P(3,2)};
            \node at (2.5, 0.5) {P(3,3)};
            \node at (3.5, 0.5) {P(3,4)};
            \node at (4.5, 0.5) {P(3,5)};
        \end{tikzpicture}
        \caption{Band-partitioning of 15 processing elements (GPUs) into a 2D logical process grid with $n_c = 5$ column groups and $n_r = 3$ row groups.}
    \label{fig:2D}
\end{figure}

The central idea is to replicate the FE discretized Hamiltonian and overlap matrices, $\bH$ and $\bM$, across
discrete groups of GPUs. To this end, the available GPUs are partitioned into a 2D logical process grid with $n_r$ rows and $n_c$ columns, where all GPUs in the same column form a column group. Figure~\ref{fig:2D} illustrates such an arrangement with $n_c = 5$ column groups and $n_r = 3$ row groups. The matrices $\bH$ and $\bM$ are replicated across column groups, with each group's copy partitioned row-wise over its $n_r$ GPUs. The trial multivectors
$\bX \in \mathbb{C}^{2M \times N_e}$ encountered during filtering are partitioned column-wise across
column groups, with each group's share further distributed row-wise among its $n_r$ constituent GPUs. Each column
group therefore operates on an independent subset of the columns of $\bX$
and evaluates the corresponding matrix--multivector products entirely within the
group, eliminating all inter-group communication during the filtering step. 
This replication is feasible because $\bH$ and $\bM$ are highly sparse, so their replicated per-GPU footprint is small in absolute terms. The dense trial multivectors $\bX$ dominate memory, but being distributed across GPUs, their per-GPU footprint also shrinks at scale.
In the limiting case of $n_r = 1$ (one GPU per column group), the
communication overhead during the filtering step is eliminated entirely, subject
to sufficient memory being available per GPU.

The subsequent Rayleigh-Ritz (RR) step, by contrast, is performed optimally with a
single column group ($n_c = 1$), corresponding to a 1D block row decomposition. The RR step as described in Algorithm~\ref{alg:RChFSI} involves evaluating projected quantities
of the form $\bY_q^{\dagger}\bH\bY_q$ and $\bY_q^{\dagger}\bM\bY_q$, yielding $N_e \times N_e$ matrices that are small in practice since $M \gg N_e$.  With $n_c > 1$, however, each column group holds only a subset of the columns of $\bY_q$, necessitating $\mathcal{O}(MN_e)$ inter-group communication to assemble the full projection, an overhead that grows with $n_c$.
The proposed band-partitioning approach therefore employs $n_c > 1$ during the filtering step, where intra-group communication can be hidden through computation-communication overlap, and reverts to $n_c = 1$ for the RR step, reconciling these two competing requirements.

\begin{algorithm}[!h]
\caption{The Rayleigh-Ritz (RR) step with band-partitioning strategy}
\label{alg:rr_bp}
\begin{algorithmic}[1]
\Require $\bY_q \in \mathbb{C}^{2M \times N_e}$ \hfill $\triangleright$ Output of the filtering step
\Require $\bH\in\mathbb{C}^{2M\times 2M}$, $\bM\in\mathbb{R}^{2M\times 2M}$
\State $\mathbf{Z_M} \leftarrow \bM\bY_q$
\State $\mathbf{Z_H} \leftarrow \bH\bY_q$
\Statex \hfill \textit{\small $\triangleright$ Steps 3--6: 2D block $\to$ 1D block row layout conversion}
\ForAll{$A \in \{\bY_q,\, \mathbf{Z_M},\, \mathbf{Z_H}\}$}
    \State $\texttt{AlltoAll}(A)$ in Row Group
    \State $\texttt{ConvertLayout}(A)$
\EndFor
\State $\bX, \bLam = \texttt{standard RR}(\bY_q, \mathbf{Z_M}, \mathbf{Z_H})$
\Statex \hfill \textit{\small $\triangleright$ Steps 8--9: 1D block row $\to$ 2D block layout conversion}
\State $\texttt{ConvertLayout}(\bX)$
\State $\texttt{AlltoAll}(\bX)$ in Row Group
\State \Return $\bX,\; \bLam$ \hfill $\triangleright$ Output eigenvectors and eigenvalues
\end{algorithmic}
\end{algorithm}

Algorithm~\ref{alg:rr_bp} details the modified RR step under the band-partitioning strategy. At the start of each RR step, the data layout of trial multivectors is converted from the 2D block
decomposition to the 1D block row decomposition via three \texttt{AlltoAll} collectives within each row group, one for each of $\bY_q$, $\mathbf{Z_M}$, and $\mathbf{Z_H}$, followed by the corresponding blockwise transposes (\texttt{ConvertLayout}). The standard RR substeps, namely \texttt{RR-Proj},
\texttt{RR-diag}, and \texttt{RR-SR}\cb{,}\cn{} are then executed with communication overhead no greater than that of the existing approach with no band partitioning.
Once complete, the inverse layout conversion is applied to
restore the 2D block decomposition before the next ChFSI iteration begins. The
optimal choice of $n_c$ therefore reflects the trade-off between the communication
savings in the filtering step and the additional \texttt{AlltoAll} overhead
incurred during the layout conversions at each RR step.

Since \texttt{AlltoAll} and \texttt{ConvertLayout} are out-of-place operations, each layout conversion requires an additional memory buffer of the same size as $\bX$ (or $\bY_q$). Due to current implementation
constraints, the overlap and the Hamiltonian matrix--multivector products ($\bM\bY_q$ and $\bH\bY_q$) each
require separate buffers, increasing the total additional memory overhead
for $4\times$ the size of $\bY_q$.
Let $s_{\mathbb{C}}$ denote the size (in bytes) of a complex scalar. The total additional memory cost per GPU is therefore $8\,s_{\mathbb{C}}\,MN_e/n_g + (n_c - 1)\,\mathcal{M}_{\bH,\bM}/n_g$, where
$n_g = n_r \times n_c$ is the total number of GPUs and
$\mathcal{M}_{\bH,\bM}$ denotes the combined memory size (in bytes) of $\bH$ and $\bM$
in the non-replicated ($n_c = 1$) case.
The second term accounts for the $(n_c - 1)$
additional copies of $\bH$ and $\bM$ introduced by the band-partitioning
replication. As detailed in Algorithm~\ref{alg:rr_bp}, each RR step requires $4\times$ \texttt{AlltoAll} collectives under the band-partitioning strategy. In the standard $\alpha$--$\beta$ latency--bandwidth model, the corresponding \texttt{AlltoAll} communication cost per layout conversion is approximately
$\alpha(n_c - 1) + \beta\,s_{\mathbb{C}}MN_e/n_g$. Both the memory cost of trial multivectors and
the bandwidth component of the communication cost scale as
$\mathcal{O}(MN_e/n_g)$ and diminish as $n_g$ increases, making
the band-partitioning strategy particularly effective in extreme-scaling regimes.

\section{Performance evaluation}
We evaluate the proposed strategies on representative benchmarks using the ALCF Aurora~\cite{allcock2025auroraarchitectingargonnesexascale} and OLCF Frontier~\cite{OLCFFrontier} exascale supercomputers.
\subsection{Systems and environment}
The ALCF Aurora supercomputer is an HPE Cray EX system comprising 10,624 compute nodes. Each node contains two 52-core Intel Xeon CPU Max Series processors and six Intel Data Center GPU Max 1550 (Ponte Vecchio) GPUs, each comprising two tiles. Eight 200-Gbps HPE Slingshot-11 NICs provide an aggregate peak network bandwidth of 1.6~Tbps per node. Performance results on Aurora reported in this section treat each separately exposed tile as one GPU device, giving 12 devices per node. The OLCF Frontier supercomputer is an HPE Cray EX system comprising 9,856 compute nodes. Each node contains one 64-core AMD Optimized 3rd Gen EPYC CPU and four AMD Instinct MI250X accelerators, each with two Graphics Compute Dies (GCDs). Four 200-Gbps HPE Slingshot-11 NICs provide an aggregate peak network bandwidth of 800~Gbps per node. Performance benchmarks on Frontier reported in this section treat each separately exposed GCD as one GPU device, giving eight devices per node. Information on the compiler and library versions is available in the AD/AE Appendix.
\subsection{Applications used to measure performance}
We evaluate our framework on two scientifically important classes of twisted bilayer systems exhibiting noncollinear magnetism and spin-orbit coupling. The first is a twisted bilayer Fe$_3$GeTe$_2$ (FGT) system, a layered van der Waals ferromagnet that has attracted intense interest for its itinerant 2D ferromagnetism~\cite{Fei2018} and gate-tunable room-temperature magnetic order~\cite{Deng2018}, making it a prime candidate for spintronic device applications. The second class consists of twisted bilayer systems of MoSe$_2$, a heavy-element transition-metal dichalcogenide in which SOC-induced band splitting, spin-valley locking, and emergent Moir\'{e} physics~\cite{Wu2017,Seyler2019} necessitate large supercells for predictive simulations. The twisted bilayer structures were generated using the Twister code~\cite{Twister}. All calculations employ SG15 optimized norm-conserving Vanderbilt pseudopotentials~\cite{oncv2013,Schlipf2015} within the PBE-GGA exchange-correlation functional~\cite{pbe}. The FE discretization uses adaptive meshes with polynomial order $p=7$, and the \emph{a priori} mesh refinement is based on each FE cell's distance from the nearest atom. For FGT, the mesh size near nuclei is 1.3~Bohr, and for MoSe$_2$ it is 2.1~Bohr. These meshes are chosen to achieve an accuracy of $10^{-4}$~Ha/atom, verified on smaller representative systems. The benchmark systems and their parameters are summarized in \cref{tab:systems}.
\begin{table}[ht]
\centering
\caption{Benchmark systems. DoFs include both spinor components.}
\label{tab:systems}
\setlength{\tabcolsep}{4pt}
\begin{tabular}{lcccc}
\hline
System & Twist angle (deg) & Atoms & Valence electrons & DoFs/atom \\
\hline
FGT              & 7.3  &   732 &  11{,}468 & $\sim$45k \\
MoSe$_2$         & 2.1  & 4{,}326 &  37{,}492 & $\sim$16k \\
MoSe$_2$         & 1.5  & 9{,}114 &  78{,}988 & $\sim$16k \\
\hline
\end{tabular}
\end{table}

\subsection{Matrix-free GPU-accelerated conjugate gradient solver}
\label{sec:results-mfcg}
\begin{figure}[!h]
\centering
\includegraphics[scale=0.65]{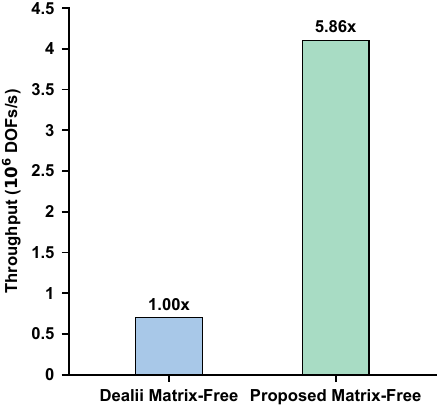}
\caption{Throughput (DoFs/s) comparison between the proposed matrix-free
$\mathbf{K}\mathbf{x}$ kernel and the deal.II~9.7.1 (Kokkos~4.7.01) implementation in the proposed preconditioned conjugate gradient solver.}%
\label{fig:matrixfree_comparison}
\end{figure}

\cref{fig:matrixfree_comparison} compares the throughput of our proposed matrix-free $\mathbf{K}\mathbf{x}$ kernel against the deal.II~9.7.1 matrix-free implementation~\cite{2025:arndt.bangerth.ea:deal}, which uses Kokkos~4.7.01 \cite{KokkosEcosystem2021} as its GPU backend. Both kernels are timed inside the same PCG solver for the 732-atom FGT system on 32 Aurora nodes (384 GPUs), where our implementation achieves $5.86\times$ the throughput in DoFs/s. The comparison is like-for-like between two sum-factorized matrix-free kernels. The \texttt{FEEvaluation} framework of deal.II~\cite{KronbichlerKormann2012,KronbichlerKormann2019} exploits the same tensor-product structure of the FE basis as our kernel, and the two runs share the same finite-element discretization ($p=7$) and the solver configuration (PCG with a degree-$5$ Chebyshev--Jacobi preconditioner and an absolute residual tolerance of $10^{-10}$). Both run entirely in FP64 without the mixed-precision or block floating-point compression approaches discussed in Section III, which are used only in the eigensolver and not for the Poisson matrix-vector product. The measured gains, therefore, reflect the kernel-level optimizations alone rather than a difference in precision or algorithm. A prior GPU-optimized matrix-free kernel with a comparable speedup was reported in~\cite{Panigrahi2024}. The present implementation introduces a register--shared memory hybrid that reduces shared memory consumption per thread block, increases thread block occupancy, and thereby enables higher FE polynomial orders that would otherwise exceed shared memory capacity.

\subsection{Low-precision computation and compressed communication enabled by R-ChFSI}
\label{sec:results-lowprec}
\cref{tab:robustness} demonstrates the robustness of R-ChFSI-enabled mixed-precision computation and compressed communication for the 732-atom FGT system on 256 Aurora nodes (3072 GPUs). All runs use an SCF stopping tolerance of $5\times10^{-4}$ based on the charge and magnetization densities. We denote the compressed formats with $n_{\mathrm{bits}} = 16$, $12$, and $10$ by BFP16, BFP12, and BFP10, respectively. The FP64/FP64 baseline converges in 38 SCF iterations and takes around 41 filtering passes aggregated over all SCF iterations. BFP16, BFP12, and BFP10 converge in 39, 38, and 39 SCF iterations and 44, 44, and 49 passes, respectively, yielding $1.84$--$2.02\times$ speedups over the 2873~s baseline with energy differences below $2\times10^{-8}$~Ha/atom.

\begin{table}[!ht]
\centering
\caption{R-ChFSI convergence for 732-atom FGT on 256 Aurora nodes. Passes are accumulated over the SCF solve; wall time excludes initialization; $\Delta E$ is relative to FP64/FP64.}
\begin{tabular}{lcccc}
\hline
Compute/Comm. & Passes & SCF & Wall time (s) & $\Delta E$ (Ha/atom) \\
\hline
FP64/FP64           & 41 & 38 & 2873 & $--$ \\
FP32(TF32)/BFP16    & 44 & 39 & 1540 & $1.9\times10^{-8}$ \\
FP32(TF32)/BFP12    & 44 & 38 & 1420 & $3.4\times10^{-9}$ \\
FP32(TF32)/BFP10    & 49 & 39 & 1565 & $7.3\times10^{-9}$ \\
\hline
\end{tabular}
\label{tab:robustness}
\end{table}

\begin{figure}[!hb]
    \centering    \includegraphics[width=0.42\textwidth]{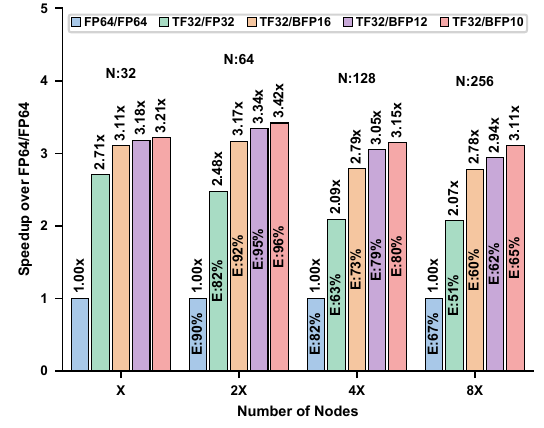}
\caption{Strong scaling of Chebyshev filtering on Aurora. Speedups are relative to FP64/FP64 at each node count $N$; Efficiency $E$ is relative to 32 nodes.}
    \label{fig:CFAurora}
\end{figure}

\begin{figure}[!ht]
    \centering
    \includegraphics[width=0.42\textwidth]{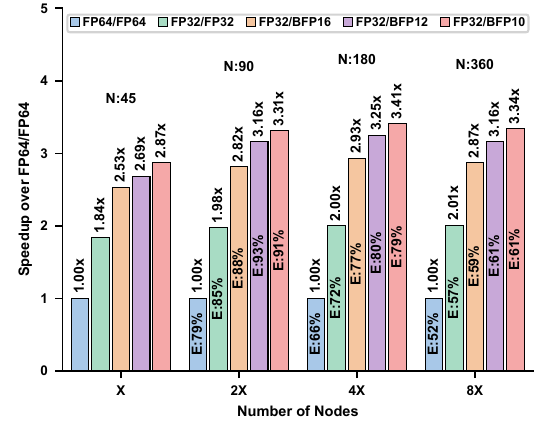}
    \caption{Strong scaling of Chebyshev filtering on Frontier. Speedups are relative to FP64/FP64 at each node count $N$; Efficiency $E$ is relative to 45 nodes.}
    \label{fig:CFFrontier}
\end{figure}

Figures~\ref{fig:CFAurora} and~\ref{fig:CFFrontier} show strong scaling of the Chebyshev filtering (CF) step from X to 8X nodes on Aurora
($32$--$256$ nodes) and Frontier ($45$--$360$ nodes). On Aurora, the
\texttt{GEMM}s use TF32 tensor cores. All speedups are relative to the FP64/FP64
baseline at the corresponding node count. All quantities
reported are computed using data from the 2nd SCF iteration onwards. At every node count on both machines,
compressed communication with $n_{\mathrm{bits}}\leq16$ delivers CF speedups of
$2.8$--$3.4\times$ on Aurora and $2.5$--$3.4\times$ on Frontier, compared with $2.1$--$2.7\times$ and $1.8$--$2.0\times$, respectively, for
FP32 communication, with BFP10 achieving the highest speedups. Scaling efficiencies are computed for each configuration
relative to its smallest node count. At 8X node count, the compressed routes
retain $60$--$65\%$ efficiency on Aurora and $59$--$61\%$ on Frontier, compared
with $51\%$ and $57\%$ for FP32 communication, respectively. On Frontier, compressed communication achieves higher scaling efficiency than
the FP64/FP64 baseline, while on Aurora, it remains comparable. The improvement is most pronounced for Aurora's TF32/FP32 configuration, which uses TF32 computation with FP32 communication, where the higher TF32 throughput makes
communication costs a larger fraction of the CF time, resulting in degraded scaling. Consequently, compressed communication
improves the scaling efficiency of TF32/FP32 by $9\%$ to $17\%$ points across all node counts. 

BFP12 provides a favorable balance between accuracy and performance across these tests and is therefore used as the default compressed-communication format in the subsequent studies.

\subsection{Band-partitioning strategy}
\begin{figure}[!ht]
    \centering    \includegraphics[width=0.42\textwidth]{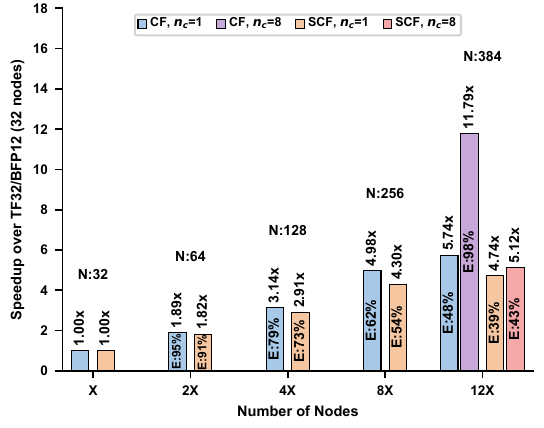}
    \caption{Band-partitioned strong scaling of CF and per-SCF time on Aurora for TF32/BFP12; speedups are relative to the 32-node $n_c=1$ run.}
    \label{fig:band-time}
\end{figure}

\begin{figure}[!ht]
    \centering    \includegraphics[width=0.42\textwidth]{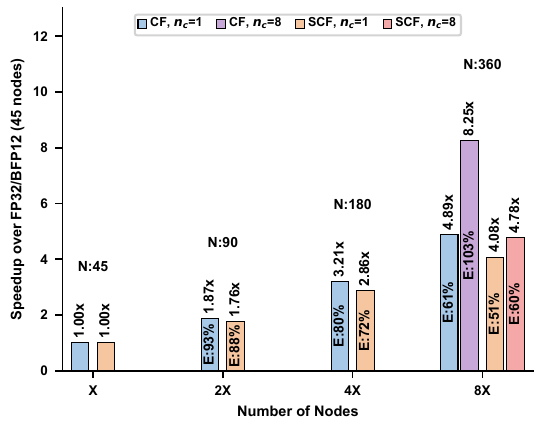}
    \caption{Band-partitioned strong scaling of CF and per-SCF time on Frontier for FP32/BFP12; speedups are relative to the 45-node $n_c=1$ run.}
    \label{fig:band-time-frontier}
\end{figure}

To improve the scaling efficiency of the Chebyshev filtering (CF) in the
extreme-scaling regime, we employ a band-partitioning strategy with $n_c = 8$
for the 384-node (12X, X=32) and the 360-node (8X, X=45) runs on the Aurora and Frontier supercomputers, respectively. Results from the 2nd SCF iteration onwards are compared
against the TF32/BFP12 and FP32/BFP12 baselines ($n_c = 1$) on Aurora and Frontier supercomputers, respectively.
Figures \ref{fig:band-time} and \ref{fig:band-time-frontier} show that, for the
baseline ($n_c = 1$), the CF scaling efficiency drops to $48\%$
at 12X on Aurora and to $61\%$ at 8X on Frontier. The band-partitioning
strategy with $n_c = 8$ raises the CF scaling efficiency to $98\%$
at 12X on Aurora and to $103\%$ at 8X on Frontier, achieving a peak CF speedup of
$11.8\times$ relative to the 32-node run on Aurora and $8.3\times$ relative to the 45-node run on Frontier. Scaling efficiency measured across the full SCF iteration also benefits. However, the growing gap between the CF and
SCF iteration efficiencies at scale suggests that other parts of the SCF iteration become the dominant bottleneck.

\subsection{Performance studies on large-scale systems}
\cref{tab:largescale} reports the per-SCF-iteration wall time breakdown for the two twisted bilayer MoSe$_2$ systems. The 4{,}326-atom system (37{,}492 electrons) was run on 256 Aurora nodes (3{,}072 GPUs), and the 9{,}114-atom system (78{,}988 electrons) on 800 Aurora nodes (9{,}600 GPUs). Wall times are reported for a representative SCF iteration (the 2nd, after initial subspace construction). For each Poisson solve in the DFT electrostatics, PCG is stopped when the absolute residual of the discretized system satisfies $\|\mathbf{K}\mathbf{V}_{\mathrm{el}}-\mathbf{f}\|_2<10^{-10}$. For the R-ChFSI eigensolver, these runs use one filtering pass per SCF iteration, with the spectral interval selected using an estimate of the largest eigenvalue. The reported second-SCF timings use a Chebyshev polynomial degree $q=60$ for both systems in R-ChFSI.
For the 37k- and 79k-electron systems studied here, the mixed-precision configuration accelerates filtering by $2.75\times$ and $2.85\times$ and reduces per-SCF wall time by $1.65\times$ and $1.56\times$, respectively. RR dominates for the larger system, where communication in the ELPA and distributed \texttt{GEMM} collectives limits its scalability.
\begin{table}[!ht]
\centering
\caption{Per-SCF wall times (s) for twisted bilayer MoSe$_2$ (CG: electrostatics/mixing; CF: filtering; RR: Rayleigh-Ritz). Total includes unlisted SCF operations.}
\label{tab:largescale}
\begin{tabular}{llcccc}
\hline
System & Config. & CG & CF & RR & Total \\
\hline
\multirow{2}{*}{\shortstack[l]{MoSe$_2$ 37k $e^-$\\(256 nodes)}}
  & FP64/FP64        & 2.8  & 165 & 134 & 315 \\
  & FP32(TF32)/BFP12 & 2.8  &  60 & 120 & 191 \\
\hdashline
\multirow{2}{*}{\shortstack[l]{MoSe$_2$ 79k $e^-$\\(800 nodes)}}
  & FP64/FP64        &  1.5 & 275 & 338 & 628 \\
  & FP32(TF32)/BFP12 &  1.8 &  97 & 290 & 403 \\
\hline
\end{tabular}
\end{table}

\section{Conclusion}
We present a GPU-accelerated finite-element framework for large-scale fully relativistic pseudopotential DFT calculations with noncollinear magnetism and spin-orbit coupling (NC-SOC-DFT) using finite-element discretization. A matrix-free kernel exploiting the tensor-product structure of the FE basis and a register–shared-memory hybrid achieves 5.86$\times$ the throughput of the state-of-the-art \texttt{deal.II} matrix-free implementation for the Poisson problem arising in DFT electrostatics. R-ChFSI-enabled mixed precision and BFP-compressed MPI communication accelerate filtering up to $3.4\times$ on Aurora and Frontier and provide a $2\times$ total SCF-solve speedup for Fe$_3$GeTe$_2$ (732 atoms) on Aurora. Band partitioning distributes bands using additional per-GPU memory under strong scaling, raising CF scaling efficiency from $48\%$ to $98\%$ on Aurora and $61\%$ to $103\%$ on Frontier. We demonstrate NC-SOC-DFT on Aurora, an exascale system, for twisted MoSe$_2$ bilayers up to 9,114 atoms and 78,988 electrons, the largest running in under 7 minutes per SCF iteration. This extends the scale of fully relativistic first-principles simulations.

The strategies are broadly applicable, but gains depend on problem size, hardware, and scaling regime. At smaller scales ($\sim$200--1000 electrons), double-precision ChFSI and simpler parallelization may suffice. On CPUs, mixed-precision gains diminish, while matrix-free Poisson and band-partitioning strategies remain applicable. R-ChFSI is largely basis-independent, requiring only Hamiltonian and overlap actions, residual evaluation, and a Rayleigh-Ritz step. Its mixed-precision filtering and blockwise Rayleigh-Ritz are therefore transferable to other real-space and plane-wave DFT codes. The Poisson kernel is specific to tensor-product FE operators, while band partitioning can be adapted to plane-wave codes, including \texttt{Quantum ESPRESSO}, through data-layout changes. Compression in FFT-based plane-wave codes would require algorithmic redesign and convergence validation owing to global transposes.

\section{Acknowledgments}
This research used the Aurora supercomputer at the Argonne Leadership Computing Facility (ALCF), a U.S. Department of Energy (DOE) Office of Science user facility at Argonne National Laboratory (ANL), and was supported by the U.S. DOE Office of Science, Advanced Scientific Computing Research Program, under Contract No. DE-AC02-06CH11357. This research also used the Frontier supercomputer at the Oak Ridge Leadership Computing Facility (OLCF), a U.S. DOE Office of Science user facility at Oak Ridge National Laboratory (ORNL), supported by the U.S. DOE Office of Science under Contract No. DE-AC05-00OR22725. N.K., G.P., and K.R. acknowledge financial support from the Prime Minister's Research Fellowship (PMRF), Ministry of Education, India. P.M. acknowledges support from the Google India Research Award 2023, the Anusandhan National Research Foundation (ANRF) ARG MATRICS grant (No. ANRF/ARGM/2025/002824/TS), and the National Quantum Mission, an initiative of DST, Govt of India. V.R.'s work is supported by the U.S. DOE Office of Science, Advanced Scientific Computing Research Program, under Contract No. DE-AC02-06CH11357. The authors acknowledge the use of AI tools (Claude and Codex) in proofreading and rewriting the text to improve flow. After using these tools, the authors reviewed and edited the content as needed and take full responsibility for the content of the publication.

\bibliography{journals.short.bib, ref.bib}

\bibliographystyle{IEEEtran}

\end{document}